\documentclass[conference]{IEEEtran}
\IEEEoverridecommandlockouts  

\usepackage{amsmath, amssymb, amsfonts}
\usepackage{graphicx}
\usepackage{textcomp}
\usepackage{xcolor}
\usepackage{cite}
\usepackage{float}
\usepackage{caption}
\usepackage{hyperref}
\usepackage[nameinlink, noabbrev]{cleveref}
\usepackage{mathtools, nccmath}
\usepackage{titlesec}
\usepackage{booktabs}
\usepackage{tabularx}
\usepackage{multirow}
\usepackage{array}
\usepackage{makecell}
\usepackage{diagbox}
\usepackage{rotating}
\usepackage{pgfplots}
\usepackage{enumitem}
\usepackage{lipsum}
\usepackage{geometry}
\usepackage{algorithm}
\usepackage{algpseudocode}

\usepackage{pifont}  

\usepackage{nomencl}

\def\BibTeX{{\rm B\kern-.05em{\sc i\kern-.025em b}\kern-.08em
    T\kern-.1667em\lower.7ex\hbox{E}\kern-.125emX}}
\begin{document}

\title{Multi-Objective Optimization Algorithms for Energy Management Systems in Microgrids: A Control Strategy Based on a PHIL System\\
}

\vspace{-0.5cm} 

\author{
    \small
    \IEEEauthorblockN{Saiful Islam}
    \IEEEauthorblockA{\textit{PhD student, Research Associate}\\
    \textit{Otto von Guericke University Magdeburg}\\
    \textit{SRH University of Applied Sciences}\\
    Berlin, Germany\\
    saiful.islam@srh.de}
    \and
    \IEEEauthorblockN{Sanaz Mostaghim}
    \IEEEauthorblockA{\textit{Faculty of Computer Science}\\
    \textit{Otto von Guericke University Magdeburg}\\
    \textit{Fraunhofer Institute for Transportation}\\
    \textit{and Infrastructure Systems IVI}\\ 
    \textit{Dresden, Germany}\\
    sanaz.mostaghim@ovgu.de}
    \and
    \IEEEauthorblockN{Michael Hartmann}
    \IEEEauthorblockA{\textit{Dean of Berlin School of Technology}\\
    \textit{SRH University of Applied Sciences}\\
    Heidelberg, Germany\\
    michael.hartmann@srh.de}
}

\maketitle

\begin{abstract}
In this research a real time power hardware in loop configuration has been implemented for an microgrid with the combination of distribution energy resources such as photovoltaic, grid tied inverter, battery, utility grid, and a diesel generator. This paper introduces an unique adaptive multi-objective optimization approach that employs weighted optimization techniques for real-time microgrid systems. The aim is to effectively balance various factors including fuel consumption, load mismatch, power quality, battery degradation, and the utilization of renewable energy sources. A real time experimental data from power hardware in loop system has been used for dynamically updating system states. The adaptive preference-based selection method are adjusted based on state of battery charging thresholds. The technique has been integrated with six technical objectives and complex constraints. This approach helps to practical microgrid decision making and optimization of dynamic energy systems. The energy management process were also able to maximize photovoltaic production where minimizing power mismatch, stabilizing battery state of charge under different condition. The research results were also compared with the baseline system without optimization techniques, and a reliable outcome was found. 

\end{abstract}


\section{Introduction}
\label{sec:Intro}
Microgrid (MG) is an innovative system but it can be challenging when it comes to operates in different modes. As an example When an MG operates in Islanded mode, it is more difficult because it only depends on the connected distributed energy resources (DERs). In such cases, a backup Diesel Generator (DG) can also be used as an alternative power supply with MG and photovoltaic (PV). In a standalone system, frequency constraint \cite{b11} is a significant challenge for what maximum PV generation must be ensured.

With a Grid-connected mode, MG can take enough supply from the grid if PV and its energy storage system (ESS) are not enough to provide the energy required for the load from the DERs. The grid-connected inverter can optimize its V-F automatically based on the required V-F  to ensure a smooth power supply\cite{b11}. During the operation of the MG, more challenge comes when the required voltage and frequency needs to be optimized \cite{b11}, \cite{b13} and corrected to ensure an optimum power flow. There are also different challenges like transient frequency deviation, system stability problems in critical load \cite{b12}, \cite{b14}, \cite{b15}, active (p) and reactive (q) power flow problems \cite{b12}, total harmonic distortion (THD) fault current \cite{b15}, uncontrolled DC link voltage for inverter operation  \cite{b17} and so on. 

Multi Objective Optimization (MOO) techniques can be used to identify the optimal DERs usage based on load to reduce any fluctuations. To solve issues related to energy supply, a combination of different DERs can help mitigate problems such as dependency only on one source, such as the utility grid, the diesel generator, etc. \cite{b28}. MOO can be formulated as a maximization or minimization problem.

MOO techniques \cite{b2} consider two or more conflicting objectives in parallel and have a set of Pareto-optimal solutions rather than a single solution. MOO algorithms are defined by an objective function vector that is minimized or maximized in terms of decision variables. 
In addition, there can be some inequality and equality constraints\cite{b27}for some multi-objective problems\cite{b9}. 
The set of all Pareto-optimal solutions is called the Pareto front or frontier\cite{b3},\cite{b4}. 

MOO techniques can be divided into evolutionary\cite{b2} and swarm-based techniques\cite{b4},\cite{b8}. Non-Dominated Sorting Genetic Algorithm-II (NSGA-II) and Non-Dominated Sorting Genetic Algorithm-III (NSGA-III) are well-known evolutionary-based techniques.

Swarm-based techniques\cite{b7},\cite{b8} are simple and robust techniques to determine the optimal solution. Multi-objective Particle Swarm Optimization (MOPSO)\cite{b4} is known as a swarm-based multi-objective technique.

\begin{figure*}[t]
\centering
\includegraphics[width=14cm]{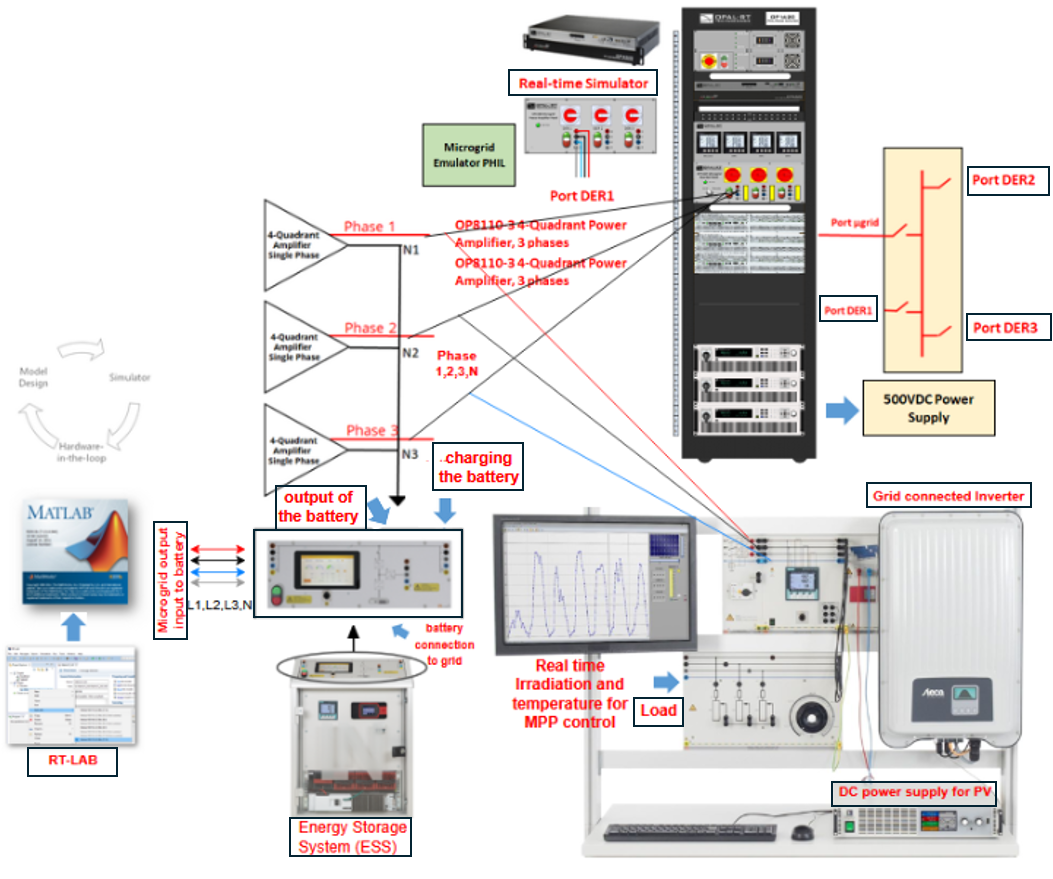}
\caption{PHIL Integration with RT Lab and Lucas Nülle test bench}
\label{fig:Fig1}
\end{figure*}

\begin{figure*}[t]
\centering
\includegraphics[width=14cm]{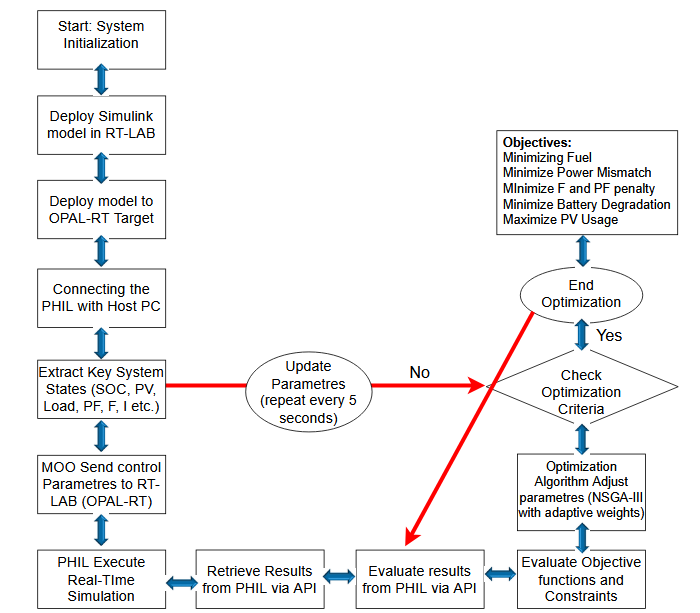}
\caption{Application Flow Diagram for Multi-Objective Optimization in PHIL Systems}
\label{fig:Fig2}
\end{figure*}

Among them, some algorithms which follow the Pareto optimal solution techniques are NSGA-II and NSGA-III\cite{b8}, the Pareto Strength Evolutionary Algorithm (SPEA), MOPSO, and Multi-Objective Evolutionary Algorithms (MOEAs)\cite{b4} are primarily used in MG. MOO techniques play an enormous role in overcoming technical, economic, and EMS challenges in an MG \cite{b5}. In order to solve technical issues using the MOO algorithm, a complex MG system has been used in this research. This paper extends a previous version currently under peer review.\footnote{A shorter version of this work has been submitted for peer review.}

The research has been divided into different sections, where section \ref{sec:Intro} explained the MG and their challenges including how optimization techniques were used in such problem statements. Section \ref{sec:Experiment} discussed about experimental setup in this research including power hardware in loop (PHIL) system. Section \ref{sec:MOO} discussed an improved MOO setup to ensure better power flow in the system while maintaining the energy management system (EMS). Section \ref{sec:Results} discuss the results and critical findings with a conclusion following section \ref{sec:Conclusion}. 

\section{Technical setup for the experiment}
\label{sec:Experiment}
To identify the real world problems a PHIL (fig.\ref{fig:Fig1}) has been tested with different components such as power amplifiers and real time simulator \cite{b6},\cite{b10}. In the experimental setup, the AC microgrid power is physically supplied by the inverter output. The DG is simulated within the real-time software environment to influence power flow decisions, such as diesel dispatch, but it does not contribute any actual physical power to the microgrid. MOO techniques has been used in terms of optimal sizing of an MG and also as real time dynamic synchronization of a system. The real time decision making was able to reduce the issues and make a stable power flow from the system. 

In this research the PHIL system has been used with a PV, grid tied inverter, battery test bench. The system configuration was integrated with a MATLAB simulation model, which has been used to activate the DC Link voltages of the amplifiers \ref{fig:Fig1}-\ref{fig:Fig2}). In this research two Power Amplifier have been used which was combined with three 500 V DC power supplies which daws 14 A currents. In the hardware configuration it was also included with a Power Amplifier Panel (OP8110) \cite{b6},\cite{b10}. The system included with a hardware in loop (HIL) based Real-Time Simulator (OP4512) which runs by the integration with tools like RT-LAB (fig.\ref{fig:Fig1}). 

In this research the PV used as a main sources of the MG system. 
In PHIL the PV can simulate upto 300 V DC to 800 V as an input for the grid tied Inverter system. It also integrated with parallel grid operation. The PV emulator also helps to simulate for different solar Irradiation condition \ref{fig:Fig1}).  
In this context of the PHIL the PV test bench acts as a DERs in the MG system. To determine the complexity of the system additional DERs such as DG, ESS and the utility grid has been added. A controllable variable load has been used to implement the MOO technique in the system \ref{fig:Fig1}). The power amplifiers (OP8110-3 and OP8110-6) have been used to provide an AC power output for grid simulation and a DC input for the PV, battery in the system. In that regard the PV test bench emulates a real solar PV system including the solar irradiation profile. The grid tied inverter has been used to converts DC generated from PV system into AC for grid connection \ref{fig:Fig1}). 

The PHIL test bench has a nominal power capacity of 5 kW with AC voltage ranges of 0 to 124 Vrms (L-N) and 0 to 240 Vrms (L-L) and supports frequencies up to 10kHz for large signals. 
Since amplifiers generate stable three-phase voltage in the MG system, the inverter was able to synchronize and feed power to the system effectively \cite{b6},\cite{b10}.

The efficiency of the PV test bench has been calculated by:
\begin{equation}
    \eta_P = \frac{P_{\mathrm{AC}}}{P_{\mathrm{DC}}}
\end{equation}
where DC power ($P_{\mathrm{DC}}$) is the power supplied by the solar generator and ($P_{\mathrm{AC}}$) is the AC converted power. The system includes also Maximum Power Point (MPP) tracking mechanism which has been also generated in real time. 

\begin{equation}
    \eta_{\mathrm{MPP}} = \frac{1}{T_M \cdot P_{\mathrm{MPP}}} \int_0^{T_M} u_{\mathrm{DC}}(t) \cdot i_{\mathrm{DC}}(t) \, dt
\end{equation}

Where $\eta_P$ is the conversion efficienc and the MPP efficiency is denoted by $\eta_{\mathrm{MPP}}$:

\begin{equation}
    \eta_{\mathrm{total}} = \eta_P \cdot \eta_{\mathrm{MPP}}
\end{equation}

\section{MOO in Energy Management System}
\label{sec:MOO}

A MG can have issues like voltage and frequency (V/F) deviation, fault in the PV strings, frequency constraints, uncertain supply and demand related problems \cite{b5}.


In this paper a MOO algorithm has been implemented to optimize the best settings for the PV power control, battery management and fuel usage minimization purposes. The optimized control parameters are used to update the DERs parameters in real time to ensure the optimized power flow in the system. Among the different MOO techniques a NSGA-III used as the core algorithm in this research. The optimizer was enabled in real-time system by responding to the dynamic inputs from a simulated MG consist of PV, ESS, DG and electrical load. The MOO algorithm used to evaluate trade-offs between conflicting operational objectives and selects a balanced decision during the live running system.

\subsection{Objective functions and constraints}

During the optimization process six practical objectives (Eqs.~(4)--(9)) has been used as fuel consumption, power mismatch, frequency (f) deviation, power factor (PF) penalty and battery degradation, and PV utilization. The system used as a fuel saving mode which integrated inside MOO optimizer. The power mismatch calculates the difference between power generation and demand.

The baseline dataset was generated using OPAL-RT, simulating real-time MG behavior without applying the MOO process. It was later compared with the MOO-based system under identical PV and load profiles. The system was initialized from a realistic state, with SOC set to 80\% and f maintained at 50.0 Hz. Importantly, the starting point did not assume maximum or minimum constraint values. The system evolved naturally based on the physical input profiles, to make realistic comparison between baseline and optimized scenarios.

The MOO optimize tries to ensure the nominal frequency which is 50 Hz. The battery degradation has been calculated by the rate of charging and discharging by means state of charge (SOC) of the battery. Each individual in the optimization represents a potential set of control decisions as follows:

\begin{align}
O_1(t) &= \text{Fuel consumption} = P_{\text{diesel}}(t) \cdot \left( \frac{\Delta t}{3600} \right) \cdot 0.4 \\
O_2(t) &= \text{Power mismatch} \nonumber \\
       &= \left| P_{\text{load}}(t) 
       - \left( P_{\text{PV}}(t) + P_{\text{diesel}}(t) + P_{\text{bat}}(t) \right) \right| \\
O_3(t) &= \text{Frequency penalty} = |f(t) - 50| \\
O_4(t) &= \text{Power factor penalty} = 1 - PF(t) \\
O_5(t) &= \text{Battery degradation penalty} = |\Delta SOC(t)| \\
O_6(t) &= \text{PV production maximization} = -P_{\text{PV}}(t)
\end{align}



\begin{align}
0 \leq P_{\text{diesel}} \leq 10 \\
-5 \leq P_{\text{battery}} \leq 5 \\
0.5 \leq P_{\text{load}} \leq 5 \\
40 \leq \text{SOC} \leq 100 \\
P_{\text{PV}}(t) \geq 0 
\end{align}

The SOC (Eqs.~\ref{eq:SOC}-\ref{eq:SOC1}) of the battery is updated at every control interval using the battery's net power and capacity. The change in SOC was calculated as:

\begin{equation}
\label{eq:SOC}
\Delta \text{SOC}(t) = \frac{P_{\text{bat}}(t) \cdot \Delta t}{C_{\text{bat}}} \times 100
\end{equation}

where \( P_{\text{bat}}(t) \) is the battery power at time \( t \) in kW, positive for charging and negative for discharging,\( \Delta t \) is the control interval duration in hours, \( C_{\text{bat}} \) is the rated capacity of the battery in kWh. The updated SOC is constrained to remain within the physical range of the battery using:

\begin{equation}
\label{eq:SOC1}
\text{SOC}(t+1) = \min \left(100, \max\left(0, \text{SOC}(t) + \Delta \text{SOC}(t)\right)\right)
\end{equation}

which ensures the SOC remains between 0\% and 100\%, representing the fully charged states of the battery. The PV output power (KW) is calculated using the DC-side measurements from the system as follows (Eqs.~\ref{eq:PV}-\ref{eq:PV1}):

where \( V_{\text{DC}}(t) \) is the measured DC voltage from the PV source at time \( t \), \( I_{\text{DC}}(t) \) is the measured DC current from the PV source at time \( t \), the resulting PV power \( P_{\text{PV}} \) is in watts (W), and typically converted to kilowatts (kW) for use in optimization.


In addition, $P_{diesel}$ represents the Power supplied by the diesel generator, $P_{bat}$ denoted as Charging/discharging power of the battery and $P_{load}$ used as the controllable load in the system. The fuel objective was modeled as a proxy proportional to diesel power output. To estimate physical fuel savings, these values were scaled by timestep duration and a nominal rate of 0.4 L/kWh.

\begin{equation}
\label{eq:PV}
P_{\text{PV}}(t) = \frac{|V_{\text{DC}}(t)| \cdot |I_{\text{DC}}(t)|}{1000}
\end{equation}

\begin{equation}
\label{eq:PV1}
PF = \frac{P_{\text{real}}}{P_{\text{apparent}}} = \frac{|P_{\text{A}}|}{V_{\text{rms}} \times I_{\text{rms}} \times 3}
\end{equation}


\begin{figure}[t]
\centering
\includegraphics[width=8cm]{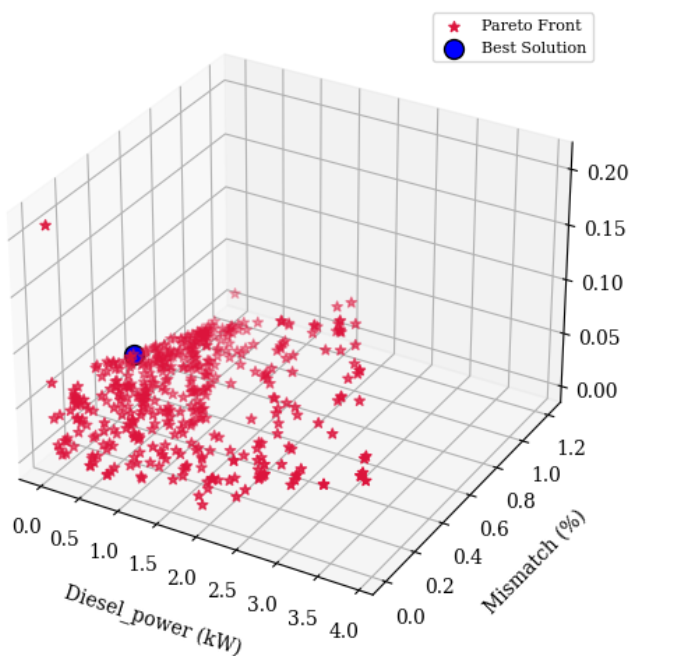}
\caption{Pareto optimal solution plot with best solution in 3D}
\label{fig:Fig3}
\end{figure}

\begin{figure}[t]
\centering
\includegraphics[width=8cm]{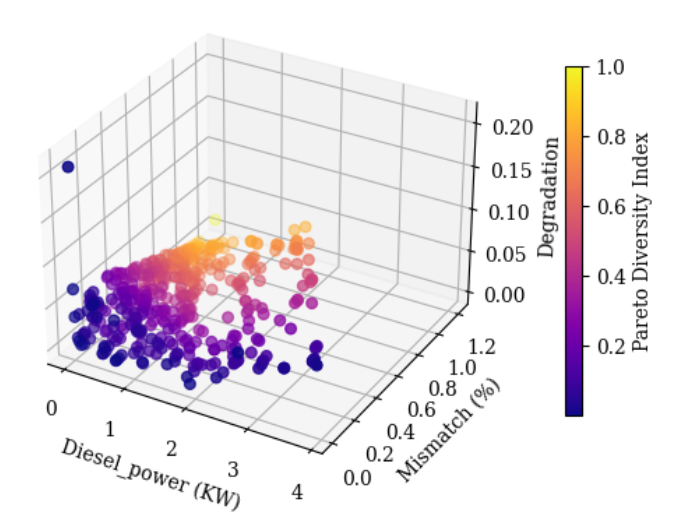}
\caption{Visualization of a 3D Pareto Front with Diversity Index Analysis}
\label{fig:Fig4}
\end{figure}

\begin{figure}[t]
\centering
\includegraphics[width=8cm]{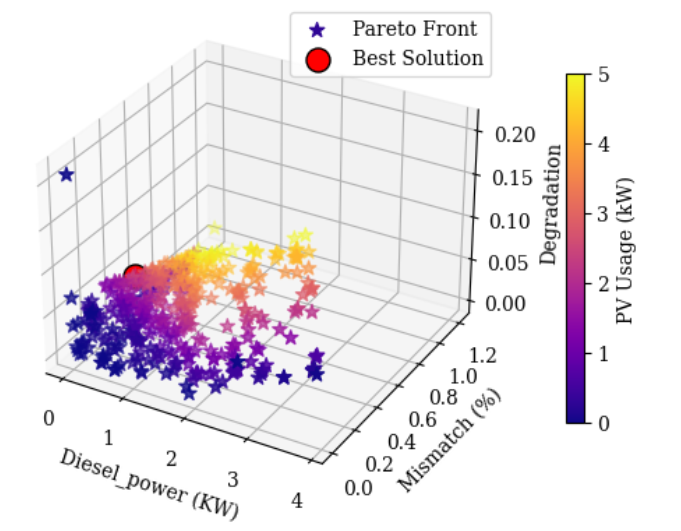}
\caption{Incorporating PV Usage as a Diversity Metric in Pareto Front Analysis}
\label{fig:Fig5}
\end{figure}

\begin{figure}[t]
\centering
\includegraphics[width=8cm]{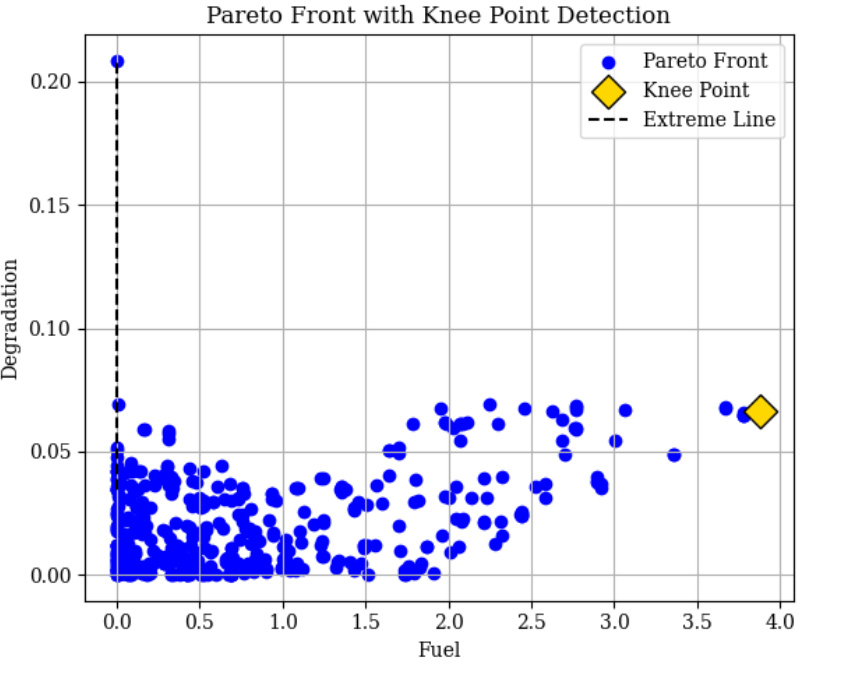}
\caption{Refining Pareto Front Analysis Through Knee Point Identification}
\label{fig:Fig6}
\end{figure}

\subsection{Energy management process}
The MOO technique used as a elitist NSGA-III algorithm which not only protect non-dominated solutions but also eliminates a few best solutions if needed to maintain a diverse pareto front using reference points \cite{b18}. NSGA-III was chosen due to its proven effectiveness in handling many-objective optimization problems. 

In terms of EMS the generated signals from the system has been streamed in real time scenario. Including MOO technique it shows more dynamic solutions by adapting the changes to DERs and loads in timestamps (Fig.\ref{fig:Fig2}). To ensure and practically find the benefits of MOO in a MG real time running system a comparison model has been also implemented between with and without MOO algorithm which is also discussed in \ref{sec:Results}. The EMS decision making process under dynamic scenario mainly focused on:

\begin{itemize}
\item A real-time dynamic MOO has been proposed which updates the system SOC after each evaluation to achieve real operation behavior during optimization. An adaptive preference-based selection method has been implemented which adjust the objective weights dynamically based on SOC level to protect it from depth of discharging (DoD).

\item The non-dominated solutions are analyses to identify the key trade-off solutions and knee points to support operational decisions. The MOO framework continuously balances fuel consumption, power quality and battery degradation and PV usage. 

\item A EMS based control strategy also used where MOO prioritize the solution that balanced PV usage maximization under different power balance conditions. Based on the observed conditions, the system prioritizes renewable energy usage when available.  

\item When PV production exceeds the load, the excess energy used to charge the battery. During the low PV production or high demand the battery discharges and supports the load. The priority of DG is lowest here to minimize the fuel cost. 
\end{itemize}

\subsection{Performance Indicators}

To evaluate the outcome of the approach a quality indicators like hyper-volume (HV), generation distance (GD) and Inverted Generational Distance (IGD) has been added in the process (Eqs.~(19)-(20)-(21)). 




\begin{algorithm}[ht]
\caption{Real-Time MOO using NSGA-III with Adaptive Weights}
\label{alg:realtime-moo}
\begin{algorithmic}[1]

\State \textbf{Initialize:} Set $\Delta t$, $C_{\text{bat}}$, SOC $\leftarrow 80\%$, $m=6$ objectives, $n=4$ variables
\State Generate reference directions $R$ (Das-Dennis); set $\mu=|R|$, generations $N$

\While{system running}
    \State Acquire OPAL data; set constraints (PV $\leq$ 5 kW)
    \State Run NSGA-III to minimize: Fuel, Mismatch, Freq Penalty, PF Penalty, Degradation, PV Usage
    \State Extract non-dominated front $F_1$
    \If{SOC $<50\%$} \State Set $w=[0.3,0.2,0.1,0.1,0.2,0.1]$
    \Else \State Set $w=[0.4,0.2,0.1,0.1,0.1,0.1]$
    \EndIf
    \For{each $i\in F_1$} \State Compute $S(i)=\sum_{k=1}^{6} w_k O_k(i)$ \EndFor
    \State \textbf{Post-Selection:} Filter solutions with Diesel/Mismatch improvement, high PV usage, acceptable Degradation
    \State Sort candidates by PV Usage $\downarrow$, Diesel Improvement $\downarrow$; select top
    \State Apply control; update SOC: $\Delta \text{SOC}=\frac{P_{\text{bat}} \Delta t}{C_{\text{bat}}}\times 100$, $\text{SOC}\leftarrow\min(100,\max(0,\text{SOC}+\Delta \text{SOC}))$
    \State Compute HV, IGD, GD; wait $\Delta t$
\EndWhile

\end{algorithmic}
\end{algorithm}

In the proposed algorithm~\ref{alg:realtime-moo} the population size has been selected as 210, and the number of generation is 200. The initial SOC taken as 80\%. The adaptive weight selection and real-control updates executed every 5 seconds. To stabilize the multi dimensional process a das-dennis method also used to generate evenly distributed reference points which guide diversity in our many-objective optimization problem with NSGA-III \cite{b22}.

In the proposed real-time multi-objective optimization framework, the Pareto front $PF$ represents the set of non-dominated solutions obtained at each control interval $\Delta t$. The optimization considers $m=6$ objectives and $n=4$ decision variables~\ref{alg:realtime-moo}. A set of reference directions $R$,with N number of generations. The battery state of charge (SOC) is initialized at $80\%$ and updated dynamically based on the extracted battery power $P_{\text{bat}}$. The SOC is recalculated using the equation $\Delta \text{SOC} = \frac{P_{\text{bat}} \times \Delta t}{C_{\text{bat}}} \times 100$, ensuring it remains within $[0, 100]\%$. To select the best solution, a weighted sum approach $\arg\min_i \sum_{k=1}^m w_k \cdot O_k(i)$ is used, where $w_k$ are adaptive weights changing according to SOC levels ~\ref{alg:realtime-moo}.

The real-time optimization process adapts the objective weighting based on the battery SOC, according to:

\[
w = 
\begin{cases} 
[0.3, 0.2, 0.1, 0.1, 0.2, 0.1] & \text{if } SOC(t) < 50\% \\
[0.4, 0.2, 0.1, 0.1, 0.1, 0.1] & \text{if } SOC(t) \geq 50\%
\end{cases}
\]

Each solution is evaluated using a weighted aggregation:

\[
\text{fitness}(i) = \sum_{k=1}^{5} w_k \cdot O_k(i)
\]

Post-optimization, performance indicators such as HV and GD are computed as (Eqs.~\ref{eq:HV}-\ref{eq:GD}-\ref{eq:IGD}):

\begin{equation}
\label{eq:HV}
HV = \int_{\text{front}} \text{dominated area}
\end{equation}

\begin{equation}
\label{eq:GD}
GD = \frac{1}{N} \sum_{i=1}^{N} \left( \text{distance}(f_i, f_{\text{ideal}}) \right)
\end{equation}

\begin{equation}
\label{eq:IGD}
\text{IGD}(A, P^*) = \frac{1}{|P^*|} \sum_{y \in P^*} \min_{x \in A} \| x - y \|
\end{equation}





\section{Results and Critical discussions}
\label{sec:Results}
From the real time dynamic PHIL system we have compared the output of system including MOO techniques with the PHIL without MOO process. It has been found that the generated solutions fully avoided the diesel usage and achieved a 3.2 kWh improvement in per control cycle. The power factor also controlled well by MOO and enhanced inverter and grid compatibility. The optimize maintains optimal behavior and confirmed robustness of the system.

The MOO-based dispatch consistently achieved low mismatch rates which is nearly 1. 05\% while maximizing the solar usage of 4.9 kW. Although the improvement in absolute mismatch over baseline was between 5–12\%, the optimizer ensured stable, renewable-prioritized dispatch with minimal diesel use and low battery degradation, even under constrained PV conditions. The power mismatch represents the deviation between total supplied power and the actual load. For each solution by:\[
M_{\text{MOO}} = \left| \frac{P_{\text{load}} - P_{\text{supply}}}{P_{\text{load}}} \right| \times 100
\] where \( P_{\text{load}} \) is the actual power demand (load), \( P_{\text{supply}} = P_{\text{PV}} + P_{\text{battery}} + P_{\text{diesel}} \) is the total power supplied. To evaluate how much the MOO solution improved mismatch compared to a baseline (non-optimized dispatch), the following improvement metric is used: \[
\text{Mismatch\_Improvement\%} = \left( \frac{M_{\text{baseline}} - M_{\text{MOO}}}{M_{\text{baseline}}} \right) \times 100
\] A higher value indicates better improvement in meeting the load accurately, with lower oversupply or under-supply.

\begin{itemize} 

\item In fig.\ref{fig:Fig3} a 3D visualization of non-dominated solutions across fuel usage, power mismatch, and battery degradation has been presented. In this research the framework of the application has been mentioned in fig.\ref{fig:Fig2}. In fig.\ref{fig:Fig3} the output results has been presented where the plot mentioned the best solutions and non-dominating solutions as a Pareto front in the picture. 

\item The fig.\ref{fig:Fig5} shows PV usage as a index including with objective values fuel consumption, power mismatch and battery degradation. Fuel consumption showed minimized as the lower diesel size (KW) has been suggested from the MOO process. The result shows the minimization of battery degradation and maximizing PV usage. The plots shows the optimization maximized PV usage, minimized diesel consumption, and reduced mismatch, while selecting the best compromise solution among many Pareto optimal candidates. The yellow color means the higher PV usage and purple means the lower usage. 

\item In the fig.\ref{fig:Fig4} a 3D visualization represents the Pareto Diversity Index, reflecting distribution richness along the front. Yellow points indicate higher diversity, supporting decision-maker flexibility. The best solution is selected based on this weighted sum from the Pareto front.

\begin{table*}[htbp]
\caption{Displaying the top optimization results from NSGA-III based on SOC}
\centering
\begin{small}
\setlength{\tabcolsep}{8pt} 
\renewcommand{\arraystretch}{1.2} 
\begin{tabular}{cccccccc}
\hline
\textbf{SOC \%} & \textbf{Mismatch \%} & \textbf{Freq} & \textbf{Battery} & \textbf{Load} & \textbf{PV Usage} & \textbf{MOO} & \textbf{Diesel} \\
 & \textbf{Improvement} & \textbf{Penalty} & \textbf{Ratio} & \textbf{Ratio} & \textbf{(kW)} & \textbf{Score} & \textbf{MOO (kW)} \\
\midrule
80 & 9.395 & 0.0 & -0.0007 & 0.9975 & 4.9686 & -0.1952 & 0.0089 \\
\midrule
80 & 7.578 & 0.0 &  0.0002 & 0.9979 & 4.9867 & -0.1935 & 0.0007 \\
\midrule
80 & 12.570 & 0.0 & 0.0321 & 0.9988 & 4.8008 & -0.1873 & 0.0131 \\
\midrule
80 & 4.752 & 0.0 & 0.0002 & 0.9999 & 4.9986 & -0.1855 & 0.0000 \\
\midrule
80 & 9.209 & 0.0 & -0.0007 & 0.9758 & 4.8648 & -0.1847 & 0.0071 \\
\midrule
80 & 9.296 & 0.0 & 0.0018 & 0.9683 & 4.8272 & -0.1828 & 0.0020 \\
\midrule
80 & 16.806 & 0.0 & 0.0847 & 0.9999 & 4.5765 & -0.1822 & 0.0000 \\
\midrule
80 & 12.011 & 0.0 & 0.0109 & 0.9994 & 4.8360 & -0.1804 & 0.0425 \\
\midrule
80 & 3.202 & 0.0 & -0.0256 & 0.9990 & 5.0000 & -0.1802 & 0.0004 \\
\midrule
80 & 17.456 & 0.0 & 0.1020 & 0.9997 & 4.4884 & -0.1754 & 0.0000 \\
\midrule
80 & 11.133 & 0.0 & 0.0614 & 0.9982 & 4.6838 & -0.1745 & 0.0000 \\
\midrule
80 & 21.229 & 0.0 & 0.1308 & 0.9969 & 4.3289 & -0.1713 & 0.0008 \\
\midrule
80 & 8.391 & 0.0 & -0.0079 & 0.9861 & 4.8577 & -0.1699 & 0.0449 \\
\midrule
80 & 12.800 & 0.0 & 0.0344 & 0.9644 & 4.6171 & -0.1697 & 0.0130 \\
\midrule
80 & 11.840 & 0.0 & 0.0109 & 0.9682 & 4.6923 & -0.1669 & 0.0375 \\
\midrule
\end{tabular}
\end{small}
\label{table:optimization_results}
\end{table*}

\item To validate our research approach, we looked at the adaptive weighted solutions alongside knee points fig.\ref{fig:Fig6}) for various objective pairs \cite{b26}. While the knee point (Fig.\ref{fig:Fig6}) indicated a balanced trade-off involving moderate diesel usage and battery degradation with a diesel power output of 3.8838 kW (equivalent to approximately 1.55L/hour of fuel consumption) (eq.4) and a degradation value of 0.0663, the adaptive weighted method (Algorithm~\ref{alg:realtime-moo}) prioritized fuel minimization and battery stress avoidance, achieving a solution with near-zero diesel use (0.0007 kW equivalent to 0.00028 L/hour). and no degradation (0.0000). This implying effective PV utilization and SOC conditions favoring renewable-driven operation. This reinforces the strength of the proposed real-time EMS approach.

\end{itemize}

From the table \ref{table:optimization_results} we can see the mismatch improvements are better than the baseline system. As the baseline system is the generated data of the MG system the PV usage was not high which has been maximized by the optimization process. It can also be seen the F deviation is nearly 0 whereas the battery ratio is also lower, which means the algorithm tries to use the maximum renewable fraction for the system. The load ratio also represents the less capacity shortage where the MOO score defines the better trade-off between all objectives by adaptive weighted methods. The PV usage was taken as negative in the objective vector because of maximization which means more PV and less use of conventional DERs.

Though the MOO technique performs well compared with the baseline PHIL system but there were some technical limitations in the system. In terms of integrating PHIL with MOO techniques, we utilized the PHIL system only upto a limit of 5 KW because of its safety regulation of DC power supply. Additionally, Physical integration of a real genset was avoided due to safety considerations and to ensure controlled experimental conditions.

\section{Conclusion}
\label{sec:Conclusion}
In this research a unique dynamic real-time adaptive MOO approach has been formulated with PHIL system. The PHIL in MG has been well known but we tried to bridge the gap between dynamic control with optimization technique. This process needs more research specially with a higher capacity for adaption. To demonstrate the outcome of the MOO in PHIL system a comparison has been also done between the real system generated data with MOO based optimized data. It has found that the system was able to make a stable SOC 80\% and reduced the power mismatch of the system. The system did not able to improve a lot f and PF as the PHIL system sizing and safety constraints does not allow to draw a higher power flow from the system. It has been also seen the usage of diesel was almost negligible and get a higher prioritization of PV. The PV usage was almost 4 KW consistently which shows a very good performance. It shows 11.93\% of improvement in power mismatch with PV capacity of 4.32 KW, 16\% improvement achieved at the higher size of 5 KW with battery degradation of 0.0183 respectively. In addition it has been also seen that SOC changes by 0.03\% in each control step which is very small and enhance battery longevity while achieving energy balance. The future work is to perform more advanced approach like hybrid MOO including surrogate models with more complex system including wind turbine and different types of inductive, capacitive or resistive loads in real time with PHIL system.

\bibliographystyle{plain} 
\bibliography{references} 

\end{document}